\documentclass[preprint,12pt]{elsarticle}

\usepackage{graphicx}

\usepackage{amssymb}
\usepackage{amsmath}

\usepackage{lineno}

\journal{Nuclear Instruments and Methods B}

\begin{document}

\begin{frontmatter}



\title{On spectral method in the axial channeling theory}


\author[Kharkov]{N.F. Shul'ga}
\author[Belgorod]{V.V. Syshchenko\corref{sysh}}
\ead{syshch@yandex.ru}
\author[Belgorod]{V.S. Neryabova}

\cortext[sysh]{Corresponding author. Tel.: +7 4722 301819; fax: +7 4722 301012}

\address[Kharkov]{A.I. Akhiezer Institute for Theoretical Physics, NSC ``KIPT'',\\ Akademicheskaya Street, 1, Kharkov 61108, Ukraine}
\address[Belgorod]{Belgorod State University, Pobedy Street, 85, Belgorod 308015, Russian Federation}

\begin{abstract}
The quantization of the transverse motion energy in the continuous potentials of atomic strings and planes can take place under passage of fast charged particles through crystals. The energy levels for electron moving in axial channeling regime in a system of parallel atomic strings (for instance, [110] strings of a silicon crystal) are found in this work for the electron energy of order of several tens of MeV, when a total number of energy levels becomes large (up to several hundreds). High resolution of the spectral method used for energy level search has been demonstrated. Hence this method could be useful for investigation of quantum chaos problem.
\end{abstract}


\end{frontmatter}



\section{Introduction}
\label{intro}
The motion of a fast charged particle in a crystal near one of crystallographic axes or planes is determined mainly by the continuous potential that is the potential of a crystal lattice averaged along the axis or plane, near which the motion takes place. The longitudinal component of the particle's momentum $p_\parallel$ parallel to the crystallographic axis or plane is conserved in such field. So, the problem on the particle's motion in a crystal is reduced to the two-dimensional problem of its motion in the transverse plane. The finite motion in the potential wells formed by the continuous potentials of atomic axes or planes is known as
axial or planar channeling, respectively (see \cite{Lindhard, Gemmel, AhSh, AhSh1, Ugg, Baz, Kum} and references therein).

The electron motion under axial channeling described by classical equation of motion can be both regular and chaotic. A pronounced example of chaotic behavior is axial channeling in a continuous potential created by two neighboring atomic strings [110] of diamond-like crystal \cite{AhSh, AhSh1}.

On the other side, the quantum effects can manifest themselves during channeling. Particularly, the quantization of the transverse motion energy can take place (see, e.g., \cite[Ch. 7, \S 53]{AhSh}, \cite[Sec. II C]{Ugg}). Investigation of the chaotic behavior on quantum level needs statistical analysis of large massive of energy levels (thousands or more) \cite{Bolotin}. Many numerical methods for searching the transverse motion energy levels as well as other quantum characteristics of a particle motion in channeling regime had been developed in the pioneering papers on quantum approach to channeling phenomenon (a good review of them could be found in the book \cite{Baz}).

The most suitable for quantum chaos studies is the so-called spectral method of energy eigenvalues searching \cite{Feit}; it was successfully used in investigation of chaotic phenomena in nuclear physics \cite{Bolotin}. In the channeling theory the spectral method had been applied for the first time in \cite{Dabagov1, Dabagov2, Dabagov3} for the case of low electron energies, $E\sim 1$ MeV, when the number of energy levels in the potential well is small. In that series of papers the evolution of incident electron wave function during penetration into a crystal was investigated, positions and widths of transverse energy bands were found as well as momentum distribution of outgoing electrons, and comparison with experimental data had been made.

Note that use of periodic crystal potential in \cite{Dabagov1, Dabagov2, Dabagov3} automatically leads to formation of band structure of the energy of transverse motion. In the paper \cite{Kozlov} high resolution of the spectral method was used to trace out how the energy levels in the potential well formed by a single atomic plane are split due to possibility of tunneling between neighboring planes. For large number of such planes this leads to formation of the band structure.

The main goal of present paper is demonstration of high resolution of the spectral method for the axial channeling case in the range of incident electron energies of order of several tens of MeV, where the number of energy levels becomes large. To simplify the problem, we consider an electron motion in the isolated two-well continuous potential formed by two [110] atomic strings of silicon crystal, neglecting the crystal periodicity.

\section{Method}
\label{method}
The spectral method of searching the energy eigenvalues of the quantum system \cite{Feit} is based on the computation of correlation function for the time dependent wave functions of the system at the initial and current time momenta, $\Psi (x,y,0)$ and $\Psi (x,y,t)$:
\begin{equation}\label{eq1}
P(t) = \int_{-\infty}^\infty \int_{-\infty}^\infty \Psi^* (x,y,0) \Psi (x,y,t) \, dxdy.
\end{equation}
Fourier transform of this correlation function,
\begin{equation}\label{eq2}
P_E = \int_{-\infty}^\infty P(t) \exp(iEt/\hbar)\, dt,
\end{equation}
contains information about the energy eigenvalues. Indeed, every solution of the time-dependent Schr\"odinger equation
\begin{equation}\label{eq3}
\hat H \Psi (x,y,t) = i\hbar \frac{\partial}{\partial t} \Psi (x,y,t)
\end{equation}
could be expressed as the superposition
\begin{equation}\label{eq4}
\Psi (x,y,t) = \sum_{n,j} A_{n,j} u_{n,j}(x,y) \exp(-iE_n t/\hbar)
\end{equation}
of the Hamiltonian's eigenfunctions $u_{n,j}(x,y)$,
$$
\hat H u_{n,j}(x,y) = E_n u_{n,j}(x,y),
$$
where the index $j$ is used to distinguish the degenerate states corresponding to the energy $E_n$. Computation of the correlation function (\ref{eq1}) for the wave function of the form (\ref{eq4}) gives
$$
P(t) = \sum_{n,n',j,j'} \exp(-iE_{n'} t/\hbar) A^*_{n,j} A_{n',j'}
$$
$$
\times \int_{-\infty}^\infty \int_{-\infty}^\infty u^*_{n,j}(x,y) u_{n',j'}(x,y) dxdy =
$$
$$
 = \sum_{n,n',j,j'} \exp(-iE_{n'} t/\hbar) A^*_{n,j} A_{n',j'} \delta_{nn'} \delta_{jj'} =
$$
\begin{equation}\label{eq5}
 = \sum_{n,j} \left| A_{n,j} \right|^2 \exp(-iE_n t/\hbar) .
\end{equation}
Fourier transformation of (\ref{eq5}) leads to the expression
\begin{equation}\label{eq6}
P_E = 2\pi\hbar \sum_{n,j} \left| A_{n,j} \right|^2 \delta (E - E_n) .
\end{equation}
We see that the Fourier transformation of the correlation function looks like a series of $\delta$-form peaks, positions of which indicate the energy eigenvalues.

So, the computation of the energy levels for the given system consists of the following steps:
\begin{enumerate}[1.]
\item Choosing the arbitrary initial wave function $\Psi (x,y,0)$. The only conditions of the choice are:
\begin{enumerate}[a)]
\item tendency to zero under $x,y\to\pm\infty$, necessary for every bound state;

\item wide spectrum that covers the depth of the potential well;

\item absence of any symmetry which could lead to the lack of some eigenfunctions in the superposition (\ref{eq4}) (see the discussion in Sec. \ref{discussion}).
\end{enumerate}
Asymmetric Gaussian waveform would be a good choice for the most cases.

\item Numerical integration of the time-dependent Schr\"odinger equation (\ref{eq3}) with the initial value $\Psi (x,y,0)$ for the discrete series of the time momenta; the value of the time step $\Delta t$ as well as other computational details are discussed in \cite{Feit, Kozlov}.

\item Computation of the integral (\ref{eq1}) for every discrete time momentum from $t=0$ to some maximal $t=T$. Subsequent integration of the obtained correlation function $P(t)$ with the exponent in (\ref{eq2}) is carried out over the finite time interval:
\begin{equation}\label{eq7}
P_E = \int_0^T P(t) \exp(iEt/\hbar)\, dt.
\end{equation}
As a result, we obtain a series of peaks of finite width (inverse proportional to $T$) instead of infinitely narrow $\delta$-like peaks (\ref{eq6}).
\end{enumerate}

Note that it is possible to find the energy eigenvalues by simple Fourier transformation of the computed wave function (\ref{eq4}) at the fixed point ($x=y=0$, for instance) as it was done in \cite{Dabagov3}. However, it is a risk to lose some eigenvalue if the particular eigenfunction $u_{n,j}(x,y)$ is equal to zero in this point.

\section{Results and discussion}
\label{discussion}
The motion of the fast charged particle in a crystal under small angle $\psi$ to the crystallographic axis densely packed with atoms could be (with good accuracy) described as a motion in the continuous string potential (e.g. the potential of the atomic string averaged along its axis) \cite{AhSh, Ugg}. The longitudinal (e.g. parallel to the string axis) component of the particle's momentum $p_\parallel$ is conserved in such a field. The motion in the transverse plane will be described in this case by the two-dimensional analog of Schr\"odinger equation \cite[Ch. 7, \S 53]{AhSh}
\begin{equation}\label{eq8}
\left\{ -\frac{\hbar^2}{2E_\parallel /c^2} \nabla^2 + U(x,y) \right\} \Psi (x,y,t) = i\hbar\frac{\partial}{\partial t} \Psi (x,y,t) ,
\end{equation}
where $\nabla^2 = \partial^2/\partial x^2 + \partial^2/\partial y^2$ is the two-dimensional Laplasian operator, and the value $E_\parallel /c^2$ (where $E_\parallel = (m^2c^4 + p_\parallel^2 c^2)^{1/2}$ ) plays the role of the particle's mass.

\begin{figure}
\begin{center}
\includegraphics[width=0.75\columnwidth]{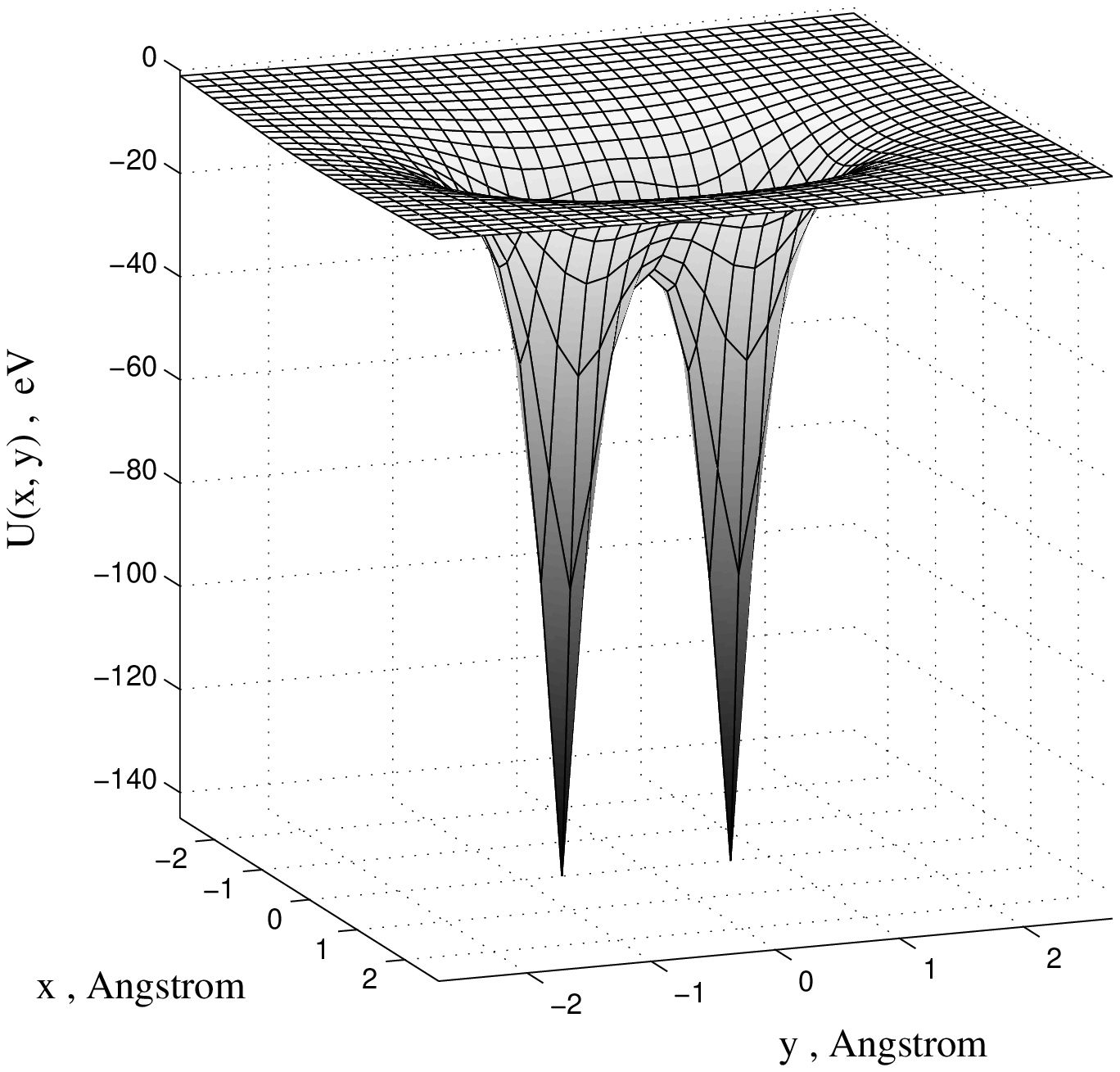}
\end{center}
\caption{Potential energy (\ref{eq10}) of an electron in the field of continuous potentials of two neighboring atomic strings [110] of a silicon crystal.}\label{Fig1}
\end{figure}

The continuous string potential could be approximated by the formula \cite[Ch. 6, \S 41]{AhSh}
\begin{equation}\label{eq9}
U_1(x,y) = - U_0 \ln \left( 1 + \frac{\beta R^2}{x^2 + y^2 + \alpha R^2} \right) ,
\end{equation}
where for the [110] string of silicon $U_0 = 60.0$ eV, $\alpha = 0.37$, $\beta = 3.5$, $R = 0.194$ \AA \ (Thomas-Fermi radius); the least distance between two parallel strings is $a/4 = 5.431/4$ \AA \ (where $a$ is the lattice period). So, the continuous potential, in which the electron's transverse motion takes place, will be described by ``two-well'' function (Fig. \ref{Fig1})
\begin{equation}\label{eq10}
U(x,y) = U_1 (x, y+a/8) + U_1 (x, y-a/8)
\end{equation}
(neglecting the influence of far-away strings). The finite motion of the electron in such potential (corresponding to negative values of the transverse motion energy $E_\perp$) is known as axial channeling \cite{AhSh}.

To search the transverse motion energy levels in the potential (\ref{eq10}) by spectral method we have chosen the initial wave function of asymmetric Gaussian form:
$$
\Psi(x,y,0) = \frac{1}{\pi\sigma_x\sigma_y} \left\{ \exp \left[ -\frac{(x-x_0)^2}{2\sigma_x^2} -\frac{(y + a/8 -y_0)^2}{2\sigma_y^2} \right] + \right.
$$
$$
\left. + \exp \left[ -\frac{(x-x_0)^2}{2\sigma_x^2} -\frac{(y - a/8 -y_0)^2}{2\sigma_y^2} \right] \right\},
$$
where $\sigma_x = 0.05$ \AA , $\sigma_y = 0.06$ \AA , $x_0 = a/35$, $y_0 = a/45$ (high symmetry of the initial wave function could lead to the absence of some eigenfunctions in the superposition (\ref{eq4}) and, as a consequence, to the loss of some energy levels, see the discussion below and the dotted curve on the lower plot on Fig. \ref{Fig3}).

\begin{figure}
\includegraphics[width=\columnwidth]{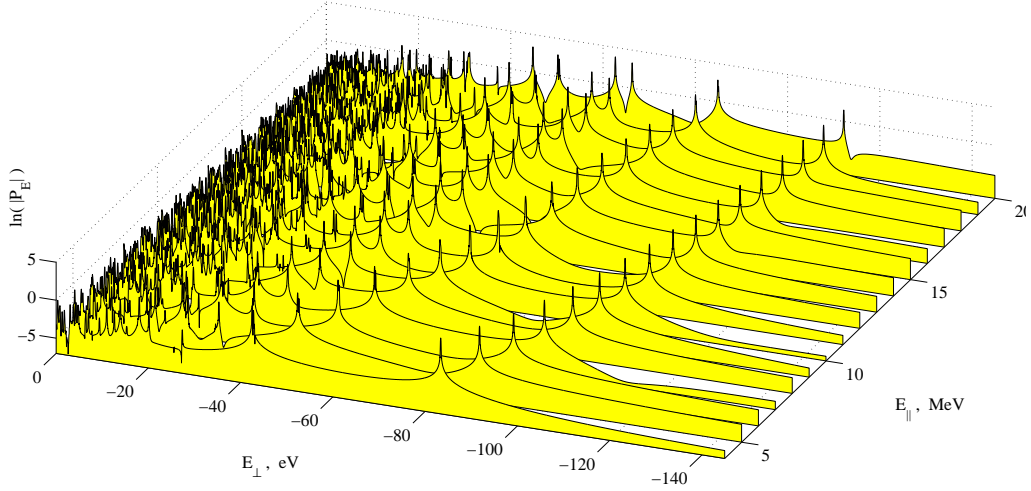}
\caption{Fourier transformations of the correlation functions (\ref{eq7}) computed for the set of values of $E_\parallel$ for an electron in the double potential well (\ref{eq10}). Positions of the maxima indicate the eigenvalues of transverse motion energy $E_\perp$.}\label{Fig2}
\end{figure}

\begin{figure}
\includegraphics[width=\columnwidth]{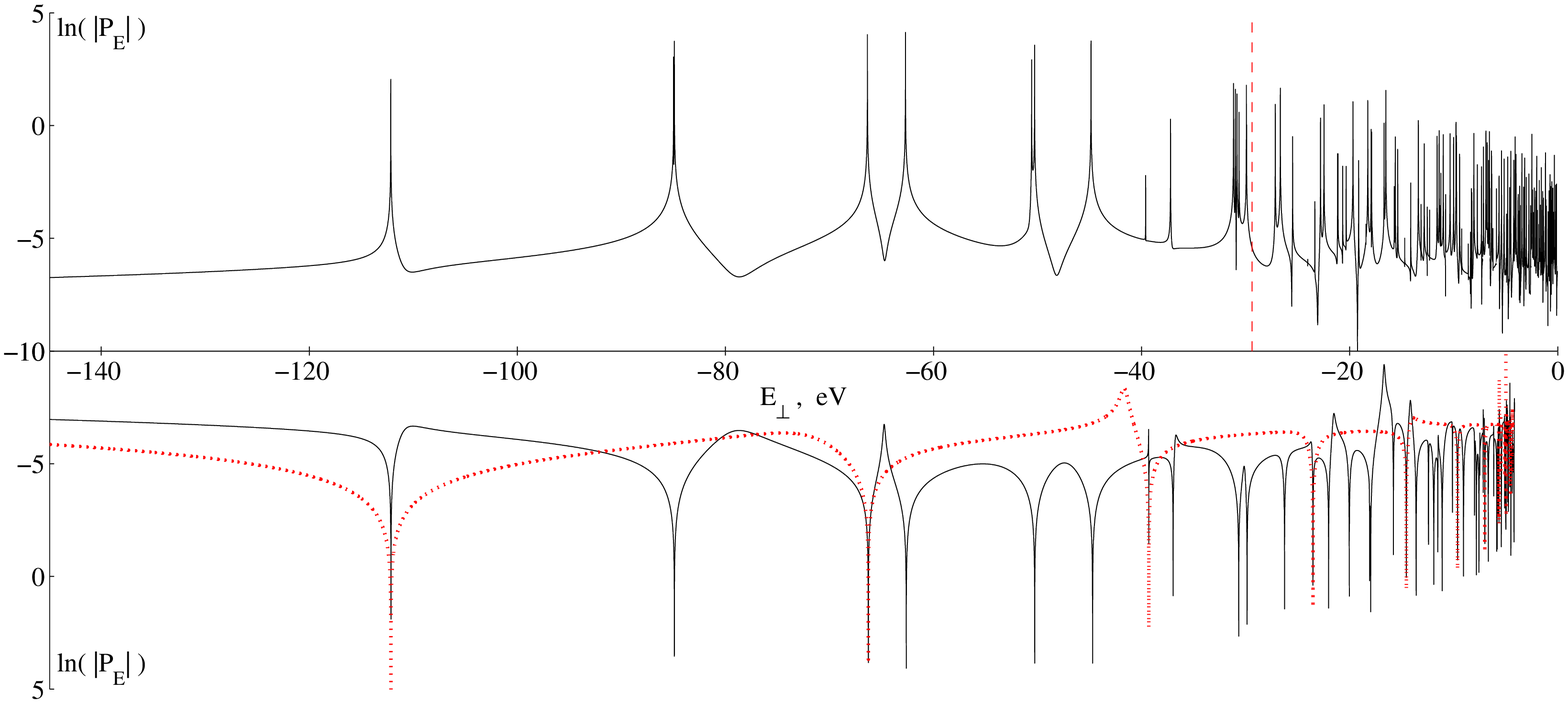}
\caption{{\it Upper plot:} Fourier transformation of the correlation function (\ref{eq7}) for the electron of energy $E_\parallel = 20$ MeV in the double potential well (\ref{eq10}) as a function of transverse motion energy $E_\perp$; vertical dashed line marks the saddle point level for the potential (\ref{eq10}). {\it Lower plot:} Fourier transformation of the correlation function for the electron in the single well (\ref{eq9}), shifted by the value $-4.11$ eV, computed for the initial wave function of general form (solid curve) and for the axially symmetric initial wave function (dotted curve).}\label{Fig3}
\end{figure}

Fourier transformations of the correlation functions for different values of $E_\parallel$ are presented in Fig. \ref{Fig2}. One can see that at the electron energy increase the transverse energy levels shift themselves deeply into the potential well, and the total number of levels in the well increases. This is the manifestation of general quantum mechanical connection between the particle mass (which role is played by the value $E_\parallel/c^2$ in our problem, as it was mentioned above) and the ground state energy as well as the total number of levels in the given potential well. The semiclassical estimation of the number of levels for the channeled electron as a function of its $E_\parallel$ could be found in \cite[Ch. 7, \S 53]{AhSh}. Some details of ``sucking-up'' of new levels into potential well from the continuum are investigated in \cite{Dabagov3}.

Results of computation of the transverse energy levels for the $E_\parallel = 20$ MeV electron in double (\ref{eq10}) and single (\ref{eq9}) potential wells are presented in Fig. \ref{Fig3}. The logarithm of the absolute value of the Fourier transform of the correlation function (\ref{eq7}) is plotted vs the $E_\perp$ value in the potential well. To simplify the comparison, the plots for both single and double wells are displayed as mirrors of each other, and the negative value $U_1(0,a/4)\approx -4.11$ eV is added to the potential energy (\ref{eq9}) to shift the bottom of a single well to the level of double one.

We see that in the range of $E_\perp$ below the saddle point of the double-well potential the general arrangement of energy levels is similar to that for a single well. However, the splitting of some levels in double-well potential is observed. For deep levels the splitting is caused by the breaking of axial symmetry for any potential well in the double-well case. As known, the quantum states in two-dimensional potential possessing axial symmetry are characterized by two quantum numbers: radial $n_r$ (that coincides with the number of zero points of the radial wave function except ones at the distances $r = 0$ and $r\to\infty$  from the center of the field) and projection $m$ of the orbital momentum to the field axis of symmetry (see, e.g., the problem 4.7 in \cite{Galitsky}). The states with $m = 0$ are non-degenerated, and the states with $|m|\not= 0$  are twice degenerated (positive and negative $m$ correspond to the same energy). Fourier transformation of the correlation functions (\ref{eq7}) for electron in a single axial symmetric potential well (\ref{eq9}) is presented by the lower plot of Fig. \ref{Fig3}. The solid curve is computed for the initial wave function of a general form, and the dotted curve is computed for axial symmetric wave function. Such special wave function contains only $m = 0$ eigenstates, hence the peaks on the dotted curve indicate only the positions of non-degenerated energy levels. The comparison with the upper plot of Fig. \ref{Fig3} demonstrates that only levels with $|m|\not= 0$ are split.

\begin{figure}
\includegraphics[width=\columnwidth]{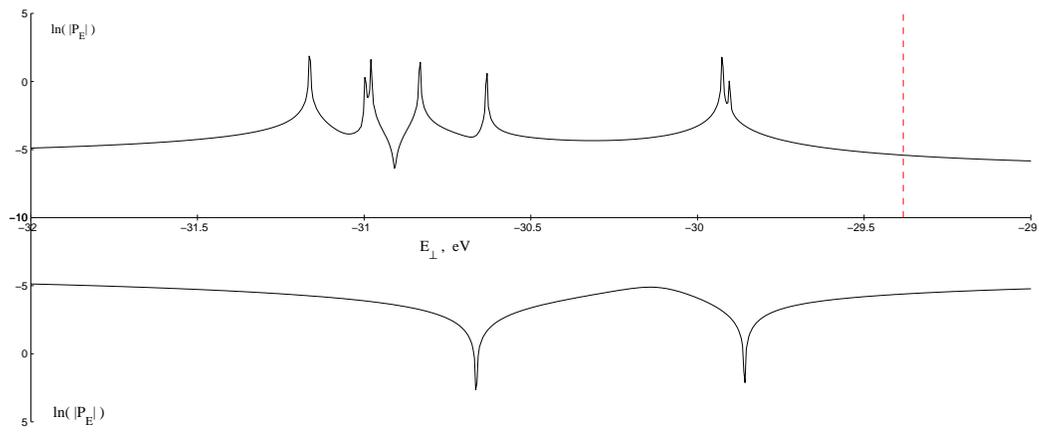}
\caption{The same as on Fig. 3, for the energy interval $-32 \leq E_\perp \leq -29$ eV.}\label{Fig4}
\end{figure}

Another mechanism of splitting is connected with the tunneling between two wells. The effect becomes visible for the levels close to the saddle point of the potential (\ref{eq10}), where the potential barrier becomes thin and easy penetrative (see Fig. \ref{Fig4}).

\section{Conclusion}
\label{conclusion}
The quantum mechanical problem on charged particle motion in oriented crystal under the axial channeling regime is considered. The spectral method of energy level search is applied to electron channeling in continuous potential of single [110] atomic string of silicon crystal as well as for duplet of such strings. The method demonstrates good resolution in the case of electron energies of order of several tens of MeV when the total number of energy levels increases up to several hundreds.

High resolution is necessary for investigation of statistical properties of large massive of energy levels; that properties could be used to study the behavior of a quantum system, which classical analog allows the dynamical chaos (axial channeling in a two-string system, see \cite[Ch. 6, \S 43]{AhSh}). The spectral method has demonstrated its effectiveness for the similar problem in nuclear physics \cite{Bolotin}.

\section{Acknowledgements}
\label{acknow}
We are grateful to Dr. V.V. Serov (Saratov State University, Russian Federation) for very helpful discussions and to Dr. A.Yu. Isupov (JINR, Dubna, Russian Federation) for the computer assistance.

This work is supported in part by internal grant of Belgorod State University, the federal program ``Academic and Teaching Staff
of Innovation Russia'', the government contract 16.740.11.0147 dated 02.09.2010, and the government contract 2.2694.2011 dated 18.01.2012.


\end{document}